\documentclass[twocolumn,reprint,amsmath,amssymb,cha,prl,aps]{revtex4-1}

\usepackage{docs}
\usepackage{epsfig}
\usepackage{graphicx}
\usepackage{bm}%
\usepackage[colorlinks=true,linkcolor=blue]{hyperref}%

\begin{document}
\title{Anomaly in the electronic structure of a BCS superconductor, ZrB$_{12}$}
\author{Sangeeta Thakur$^1$}
\author{Deepnarayan Biswas$^1$}
\author{Nishaina Sahadev$^1$}

\author{P. K. Biswas$^2$}
\altaffiliation{Present address: Paul Scherrer Institut, CH-5232
Villigen PSI, SWITZERLAND}
\author{G. Balakrishnan$^2$}

\author{Kalobaran Maiti$^1$}
\altaffiliation{Corresponding author: kbmaiti@tifr.res.in}

\affiliation{$^1$Department of Condensed Matter Physics and
Materials' Science, Tata Institute of Fundamental Research, Colaba,
Mumbai - 400 005, India}

\affiliation{$^2$Department of Physics, University of Warwick,
Coventry, CV4 7AL, UK}

\date{\today}

\begin{abstract}

We investigate the electronic structure of a complex conventional
superconductor, ZrB$_{12}$ employing high resolution photoemission
spectroscopy and {\it ab initio} band structure calculations. The
experimental valence band spectra could be described reasonably well
within the local density approximation. Energy bands close to the
Fermi level possess $t_{2g}$ symmetry and the Fermi level is found
to be in the proximity of quantum fluctuation regime. The spectral
lineshape in the high resolution spectra is complex exhibiting
signature of a deviation from Fermi liquid behavior. A dip at the
Fermi level emerges above the superconducting transition temperature
that gradually grows with the decrease in temperature. The spectral
simulation of the dip and spectral lineshape based on a
phenomenological self energy suggests finite electron pair lifetime
and a pseudogap above the superconducting transition temperature.

\end{abstract}

\maketitle
%


The relationship between pseudogap phase and superconductivity in
high temperature superconductors is an issue of discussion for many
years now \cite{Shen,Statt,Damascelli}. While one school believes
that the pseudogap arises due to some hidden order and/or effects
not associated to superconductivity, the other school attributes
pseudogap to the electron pair formation above the transition
temperature, $T_c$ as a precursor to the superconducting gap that
leads to superconductivity upon establishment of the coherence among
the electron pairs at $T_c$. Independent of these differences, it is
realized that electron correlation plays a significant role in the
occurrence of pseudogap phase in these unconventional
superconductors \cite{Superconductor,Superconductor1}. Several
questions are being asked on whether the pseudogap phase is specific
to unconventional superconductors, electron correlation or low
dimensionality is a necessity for such phase \emph{etc}.

\begin{figure}[b]
 \vspace{-6ex}
\includegraphics[scale=0.4]{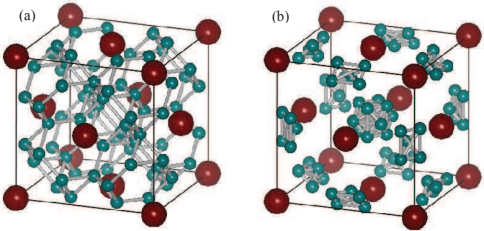}
 \vspace{-52ex}
 \caption{(a) Crystal structure of ZrB$_{12}$. (b) B$_{12}$ units are
contracted to visualize the isolated B$_{12}$ clusters with better
clarity.}
 \vspace{-2ex}
\end{figure}

Apart from these issues, the quest of new superconducting compounds
resulted into the discovery of new materials such as cubic hexane,
$M$B$_6$ \cite{cab6} and dodecaborides, $M$B$_{12}$ ($M$ = Sc. Y,
Zr, La, Lu, Th) those attracted much attention due to their
interesting electronic properties \cite{Matthias}. ZrB$_{12}$ is one
such compound exhibiting relatively high superconducting transition
temperature of 6 K in the MB$_{12}$ family \cite{Matthias}. It forms
in cubic structure as shown in Fig. 1. In Fig. 1(a), the real
lattice positions are shown with the correct bond lengths and atom
positions. Although, the structure appears to be complex, it is
essentially a rock salt structure constituted by two interpenetrated
fcc lattices formed by Zr and B$_{12}$ units as demonstrated in Fig.
1(b) by compressing the B$_{12}$ units for clarity. Various
transport and magnetic measurements suggest a
Bardeen-Cooper-Schriefer (BCS) type superconductivity in this
material, which is termed as conventional superconductivity. Since,
Zr atoms are located within the huge octahedral void space formed by
the B$_{12}$ units, the superconductivity in these materials could
conveniently be explained by the electron-phonon coupling mediated
phenomena involving primarily low energy phonon modes associated
with the vibration of Zr atoms \cite{Filippov1}.

The electronic properties of ZrB$_{12}$ manifest plethora of
conflicts involving the mechanism of the transition. For example,
there are conflicts on whether it is a type I or type II
superconductor \cite{Filippov3}, whether it possesses single gap
\cite{Daghero} or multiple gaps \cite{Gasparov2}, {\it etc.}
Interestingly, Gasparov {\it et al}. \cite{Gasparov2} found that
superfluid density of ZrB$_{12}$ exhibits unconventional temperature
dependence with pronounced shoulder at $T/T_c \sim$ 0.65. In
addition, they find different superconducting gap and transition
temperatures for different energy bands, and the values of the order
parameter obtained for $p$ and $d$ bands are 2.81 and 6.44,
respectively. Detailed magnetic measurements reveal signature of
Meissner, mixed and intermediate states at different temperatures
and magnetic fields \cite{biswas_thesis}. Some of these variances
were attributed to surface-bulk differences in the electronic
structure \cite{jap}, the superconductivity at the sample surface
\cite{Tsindlekht1,Filippov2,Tsindlekht2} \emph{etc}. Evidently, the
properties of ZrB$_{12}$ is complex despite exhibiting signatures of
BCS type superconductivity. Here, we probed the evolution of the
electronic structure of ZrB$_{12}$ employing high resolution
photoelectron spectroscopy. The experimental results exhibit
spectral evolution anomalous to conventional type superconductor and
signature of pseudogap prior to the onset of superconductivity.

\section*{RESULTS}

\begin{figure}
\vspace{-3ex}
\includegraphics[scale=0.45]{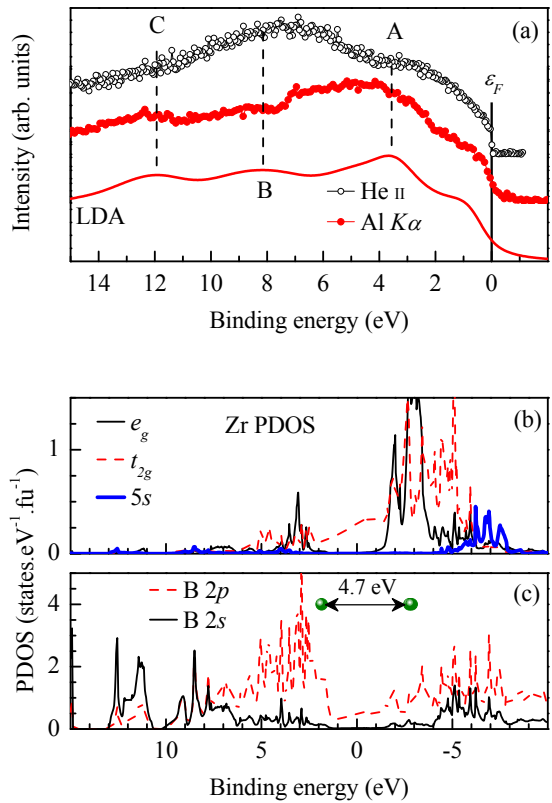}
 \vspace{-6ex}
 \caption{(a) XP and He {\scriptsize II} valence band spectra of ZrB$_{12}$.
The lines show the simulated XP spectrum from the band structure
results. Calculated partial density of states of (b) Zr 4$d$
$t_{2g}$ (dashed line) \& $e_g$ (solid line) symmetries, and (c) B
2$s$ (solid line) and 2$p$ (dashed line) states.}
 \vspace{-2ex}
\end{figure}

Valence band spectra obtained using He {\scriptsize II} and Al
$K\alpha$ excitation energies are shown in Fig. 2(a) exhibiting
three distinct features marked by A, B and C beyond 2 eV binding
energies. The intensities between 6-10 eV binding energies
represented by B are prominent in the He {\scriptsize II} spectrum,
while the relative intensity between 3-4 eV binding energies
represented by A is enhanced in the $x$-ray photoemission (XP)
spectrum. Such change in intensity in the angle integrated spectra
may be attributed to the matrix elements associated to different
constituent states forming the eigenstates of the valence band,
which is a sensitive function of the photon energy
\cite{Yeh,Yeh1,Yeh2}. Therefore, the photoemission cross section
will vary with the photon energy that can be used to identify the
orbital character of the energy bands. Considering this feature of
the technique, the spectral feature, B can be attributed to dominant
B 2$p$ orbital character and the feature A to Zr 4$d$ orbital
character.

Energy band structure of ZrB$_{12}$ has been calculated within the
local density approximations (LDA). The calculated partial density
of states (PDOS) corresponding to Zr 4$d$ \& 5$s$, and B 2$s$ \&
2$p$ states are shown in Fig. 2(b) and 2(c), respectively. Zr 5$s$
contribution appears in the energy range of 5 - 7.5 eV above the
Fermi level with negligible contributions in the energy window
studied here as shown by thick solid line in Fig. 2(b). B 2$s$
contributions appear beyond 6 eV binding energies. Dominant
contribution from B 2$p$ PDOS appears between 1.5 to 10 eV binding
energies. The center of mass of the B 2$p$ PDOS (considering both
occupied and unoccupied parts) appears around 1.9 eV binding energy.
The center of mass of the entire Zr 4$d$ PDOS appear around 2.8 eV
above $\epsilon_F$, which is about 4.7 eV above the center of mass
of the B 2$p$ PDOS. Thus, an estimate of the charge transfer energy
could be about 4.7 eV in this system. The ground state is metallic
with flat density of states across $\epsilon_F$ arising due to
highly dispersive energy bands in this energy range. The density of
states at $\epsilon_F$ possess close to 2:1 intensity ratio of the B
2$p$ and Zr 4$d$ orbital character.

\begin{figure}
 \vspace{-3ex}
\includegraphics[scale=0.45]{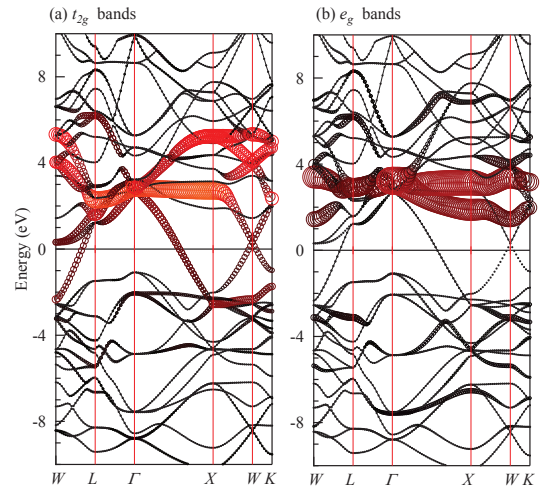}
 \vspace{-36ex}
 \caption{Energy band dispersions. Symbol size in the left panel
show $t_{2g}$ contributions and that on the right panel shows $e_g$
contributions.}
 \vspace{-2ex}
\end{figure}

The electronic states with $t_{2g}$ and $e_g$ symmetries, shown in
Fig. 2(b), exhibit almost overlapping center of mass suggesting
negligible crystal field splitting (crystal field splitting $\sim$
100 meV). The energy bands are shown in Fig. 3. $e_g$ bands are
relatively narrow and exhibit a gap at $\epsilon_F$. The intensity
at $\epsilon_F$ essentially have highly dispersive $t_{2g}$
symmetry.

Using the above band structure results, we calculated the XP
spectrum considering the photoemission cross section of the
constituent partial density of states as discussed later in the
Method section. The calculated spectrum shown by solid line in Fig.
2(a) exhibits an excellent representation of the experimental XP
spectrum. It is clear that the feature, C in the energy range 10-14
eV corresponds to photoemission signal from B 2$s$ levels. The
features, A and B arise due to hybridized B 2$p$ - Zr 4$d$ states.
The relative contribution from Zr 4$d$ states is higher for the
feature A.

\begin{figure}
 \vspace{-3ex}
\includegraphics[scale=0.45]{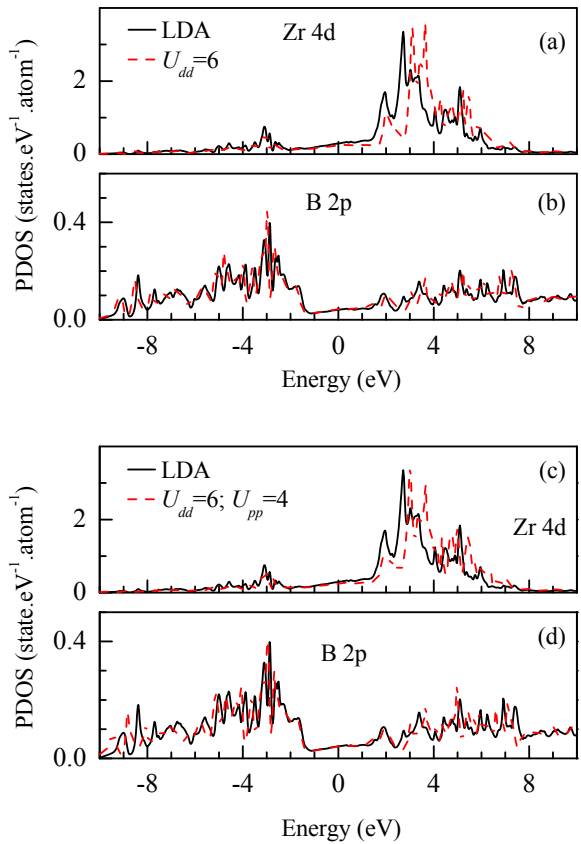}
 \vspace{-4ex}
 \caption{ (a) Zr 4$d$ and (b) B 2$p$ PDOS obtained from LDA and LDA+$U$
calculations ($U_{dd}$ = 6 eV). (c) Zr 4$d$ and (d) B 2$p$ PDOS
obtained $U_{dd}$ = 6 eV and $U_{pp}$ = 4 eV.}
 \vspace{-2ex}
\end{figure}

In order to probe the electron correlation induced effect on the
electronic structure, we have carried out electron density of states
calculation using finite electron correlation among the Zr 4$d$
electrons, $U_{dd}$ as well as that among B 2$p$ electrons,
$U_{pp}$. In Figs. 4(a) and 4(b), we show the calculated results for
Zr 4$d$ and B 2$p$ PDOS with $U_{dd}$ = 0 (solid line) and 6 eV
(dashed line) for the Zr 4$d$ electrons. Even the large value of
$U_{dd}$ of 6 eV does not have significant influence on the occupied
part of the electronic structure, while an energy shift is observed
in the unoccupied part. Such scenario is not unusual considering the
large radial extension of 4$d$ orbitals and poor occupancy
\cite{ruth}. The inclusion of an electron correlation, $U_{pp}$ of
upto 4 eV for B 2$p$ states exhibits insignificant change in the
results. Considering Zr being a heavy element, we carried out
similar calculations including spin-orbit interactions and did not
find significant modification in the results - we have not shown
these later results to maintain clarity in the figure. All these
theoretical results suggest that consideration of electron
correlation within the LSDA+$U$ method has insignificant influence
in the occupied part of the electronic structure of this system.

\begin{figure}
 \vspace{-3ex}
\includegraphics[scale=0.45]{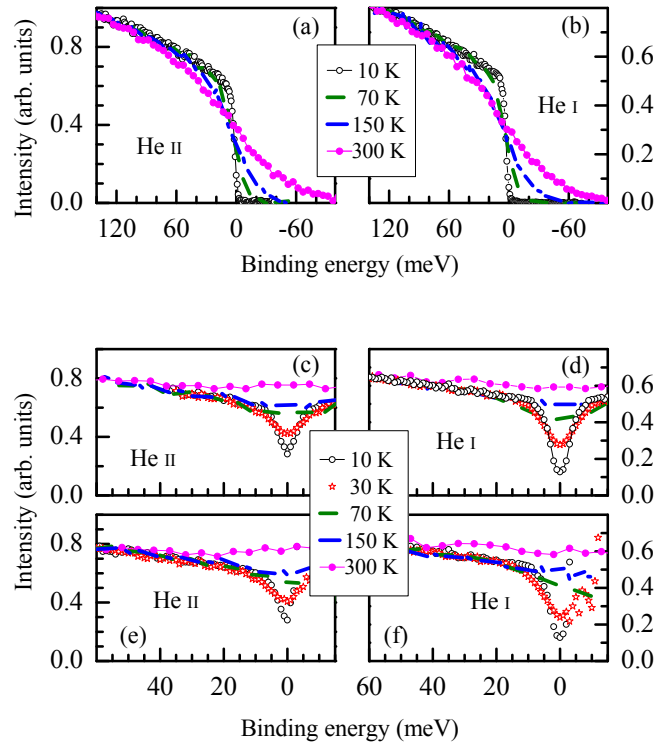}
 \vspace{-12ex}
 \caption{High resolution spectra near the Fermi level using (a) He
{\scriptsize II} and (b) He {\scriptsize I} photon energies. The
symmetrized spectral density of states (SDOS) for (c) He
{\scriptsize II} and (d) He {\scriptsize I} spectra, and SDOS
obtained by division of resolution broadened Fermi function for (e)
He {\scriptsize II} and (f) for He {\scriptsize I} spectra.}
 \vspace{-2ex}
\end{figure}

We now investigate the spectral changes near $\epsilon_F$ with high
energy resolution in Fig. 5. The spectral density of states (SDOS)
are calculated by symmetrizing the experimental spectra; $SDOS =
I(\epsilon - \epsilon_F)+I(\epsilon_F - \epsilon)$ and shown in
Figs. 5(c) and 5(d). Such an estimation of SDOS is sensitive to the
definition of the Fermi level, which is carefully derived at each
temperature by the Fermi cutoff in the valence band spectra for
silver mounted on the sample holder in electrical connection with
the sample and measured using identical experimental conditions. The
other effect is the asymmetry across the Fermi level. This can be
addressed by obtaining the SDOS by the division of the resolution
broadened Fermi-Dirac distribution function. The results are shown
in Figs. 5(e) and 5(f) for He {\scriptsize II} and He {\scriptsize
I} spectra, respectively. Clearly, the SDOS obtained following the
later procedure exhibits symmetric SDOS across $\epsilon_F$. The 30
K and 10 K data exhibit large noise above $\epsilon_F$ due to the
division of negligibly small intensities by small numbers arising
from Fermi-Dirac function at these temperatures. These results
demonstrate the reliability of SDOS extraction procedures and also
indicate signature of particle-hole symmetry in this system.

\begin{figure}
 \vspace{-3ex}
\includegraphics[scale=0.4]{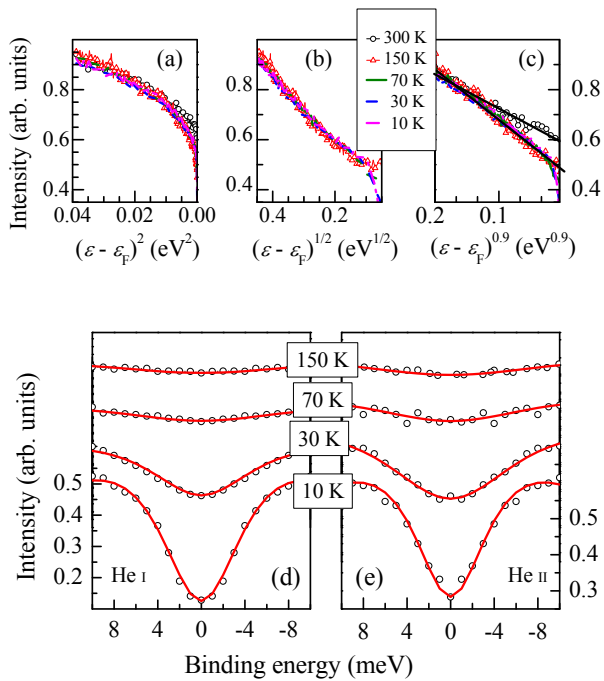}
 \vspace{-8ex}
 \caption{High resolution spectral density of states is shown as a
function of (a) $|\epsilon - \epsilon_F|^2$, (b) $|\epsilon -
\epsilon_F|^{1/2}$, and (c) $|\epsilon - \epsilon_F|^{0.9}$.
Spectral density of states from (d) He {\scriptsize I} and (e) He
{\scriptsize II} spectra. The lines represent the spectral functions
simulated to fit the experimental spectra.}
 \vspace{-2ex}
\end{figure}

The spectral lineshape near the Fermi level often provides important
information about the thermodynamic properties of the systems. For
example, a lineshape dependence of $|\epsilon - \epsilon_F|^2$
corresponds to Fermi liquid behavior \cite{Shklovskii,Shklovskii1} -
absence of linearity of SDOS with $|\epsilon - \epsilon_F|^2$ shown
in Fig. 6(a) indicates a deviation from such a behavior in the
present case. The decrease in intensity at $\epsilon_F$ with the
decrease in temperature may arise due to the disorder induced
localization of the electronic states. Altshuler \& Aronov showed
that the charge disorder in a correlated system leads to $|\epsilon
- \epsilon_F|^{0.5}$ dependence of the density of states near Fermi
level \cite{Altshuler}, which has been proved employing
photoemission spectroscopy \cite{Sarma,Fujimori}. In the present
case, a plot of SDOS with $|\epsilon-\epsilon_{F}|^{0.5}$ shown in
Fig. 6(b) exhibits deviation from linearity.

Interestingly, the curvatures at the exponents of 0.5 and 2 are
opposite indicating an intermediate exponent for the present system.
The simulation of SDOS as a function of
$|(\epsilon-\epsilon_{f})|^{\alpha}$ with $\alpha \approx 0.9$
exhibits linear dependence with energy for all the temperatures
studied (see Fig. 6(c)). Such a behavior in a conventional
superconductor is curious. Most interestingly, in Fig. 6(c), all the
data below 150 K superimpose exactly and correspond to the same
energy scale, while the data at 300 K exhibit similar energy
dependence with a different slope. These results suggest signature
of an incipient phase transition at an intermediate temperature.

The SDOS at $\epsilon_F$ is intense and flat at 300 K with large
dispersion of the bands and correspond to a good metallic phase
consistent with the band structure results. The decrease in
temperature introduces a dip at $\epsilon_F$. The dip gradually
becomes more prominent with the decrease in temperature. The same
scenario is observed in both, He {\scriptsize I} and He {\scriptsize
II} spectra indicating this behavior to be independent of the photon
energy used. This dip may be attributed to a pseudogap appearing at
low temperatures \cite{lab6}. Although the probed lowest temperature
is slightly above $T_c$, the dip grows monotonically with the
decrease in temperature - a large suppression of the spectral weight
at the Fermi level is observed at 10 K indicating relation of this
dip to the superconducting gap. In order to estimate the energy gap
in these spectral functions, we used a phenomenological self-energy,
$\Sigma_k(\epsilon)$ following the literature \cite{Randeria}.
$$\Sigma_k(\epsilon) = -i\Gamma_1 + \Delta^2/[(\epsilon + i\Gamma_0)
+\epsilon_k]$$ where $\Delta$ represents the gap size. The first
term is the energy independent single particle scattering rate and
the second term is the BCS self energy. $\Gamma_0$ represents the
inverse electron pair lifetime. The spectral functions, the
imaginary part of the Green's functions, were calculated as follows.
$A(\epsilon)=(1/\pi) Im \sum_k G_k(\epsilon) = (1/\pi)\sum_k
\Sigma_k^{\prime\prime}(\epsilon) / [(\epsilon-\epsilon_k-
\Sigma_k^\prime(\epsilon))^2 +
(\Sigma_k^{\prime\prime}(\epsilon))^2]$, where
$\Sigma_k^\prime(\epsilon)$ and $\Sigma_k^{\prime\prime}(\epsilon)$
are the real and imaginary part of the self energy. We find that a
finite value of $\Gamma_0$ is necessary to achieve reasonable
representation of the experimental data. The simulated and
experimental spectral density of states are shown by superimposing
them in Figs. 6(d) and 6(e) for the He {\scriptsize I} and He
{\scriptsize II} spectra, respectively. Evidently, the simulated
spectra provide a remarkable representation of the experimental
spectra.

\section*{Discussion}


The experimental and calculated band structure results suggest
significant hybridization between Zr 4$d$ and B 2$p$ states leading
to similar energy distribution of the PDOS. Here, the eigenstates
are primarily constituted by the linear combination of the Zr 4$d$
and B 2$p$ states. The antibonding bands possess large Zr 4$d$
orbital character and appear in the unoccupied part of the
electronic structure \cite{bndstr1,bndstr2}. The energy bands below
$\epsilon_F$ forming the valence band are the bonding eigenstates
consisting of dominant 2$p$ orbital character. The covalency between
these states could be attributed to the large radial extension of
the 4$d$ orbitals overlapping strongly with the neighboring states
as also observed in other 4$d$ systems
\cite{covalency,covalency1,covalency2}. The calculated results
within the local density approximation are consistent with the
experimental spectra - subtle differences between the experiment and
theory could not be captured via inclusion of electron correlation
and/or consideration of spin-orbit coupling in these LDA
calculations.

The band structure results exhibit two important features - (i) the
energy bands with $t_{2g}$ symmetry are highly dispersive compared
to $e_g$ bands and (ii) the crystal field splitting is negligible.
As shown in Fig. 1, the B$_{12}$ dodecahedrons form an octahedra
around Zr sites with the center of mass of the B$_{12}$ units at the
edge centers of the unit cell. The size of the dodecahedrons are
quite large with no boron situating on the unit cell edge. Thus, the
overlap of B 2$p$ orbitals with the $t_{2g}$ orbitals is
significantly enhanced due to the proximity of borons along the face
diagonals, while that with the $e_g$ orbitals is reduced leading to
a comparable crystal field potential on both the orbitals. The
larger hybridization with the $t_{2g}$ orbitals are also manifested
by the large dispersion of the $t_{2g}$ bands spanning across the
Fermi level while $e_g$ bands are relatively more localized. From
these calculations, the effective valency of Zr was found to vary
between (+2.4) to (+2.5) depending on the electron interaction
parameters such as correlation strength, spin orbit coupling
\emph{etc}.

\begin{figure}
 \vspace{-3ex}
\includegraphics[scale=0.45]{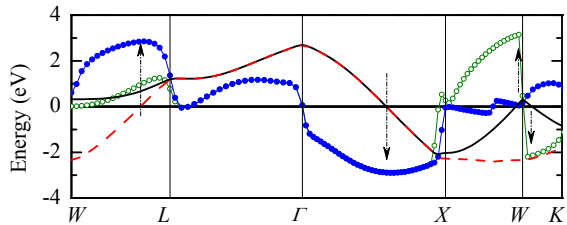}
 \vspace{-64ex}
 \caption{Energy band dispersions of the $t_{2g}$ bands crossing the
Fermi level are shown by lines. The symbols represent the first
derivative of the same energy bands. The arrows show the Fermi level
crossings.}
 \vspace{-2ex}
\end{figure}

The Fermi level is located almost at the inflection points of each
of the energy band crossing $\epsilon_F$ as shown in Fig. 7. This is
verified by investigating the first derivative of the energy bands
shown by the symbols in the figure. The arrows in the figure
indicate Fermi level crossing and corresponding point in the first
derivative curves. Evidently, the band crossing correspond to an
extremum in the derivative plot. The inflection point is the point,
where the curve changes its curvature that corresponds to a change
in the character of the conduction electrons (the Fermi surface)
from hole-like to electron-like behavior. This behavior is termed as
Lifschitz transition \cite{lifschitz}. Thus, this material appear to
be lying in the proximity of the Lifschitz transition as also
observed in various exotic unconventional superconductors such as
Fe-pnictides \cite{liu-nphys}. Due to the lack of cleavage plane and
hardness of the sample, the Fermi surface mapping of this sample
could not be carried out to verify this theoretical prediction
experimentally. It is well known that the band structure
calculations captures the features in the electronic structure well
in the weakly correlated systems as also found in the present case
suggesting possibility of such interesting phenomena in this system.
We hope, future studies would help to enlighten this issue further.

\begin{figure}
 \vspace{-3ex}
\includegraphics[scale=0.45]{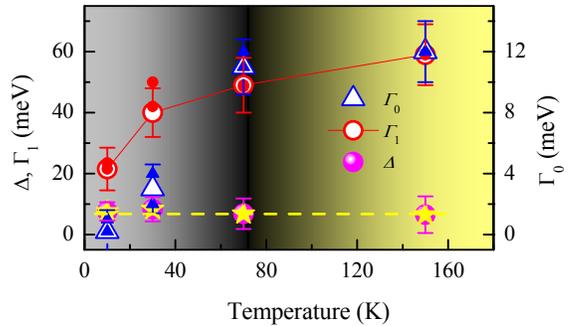}
 \vspace{-56ex}
 \caption{The gap, $\Delta$, inverse pair lifetime, $\Gamma_0$ and single particle
scattering rate, $\Gamma_1$ as a function of temperature. The open
and close symbols correspond to the fitting of He {\scriptsize I}
and He {\scriptsize II} data, respectively.}
 \vspace{-2ex}
\end{figure}

The high resolution spectra close to the Fermi level indeed exhibit
deviation from Fermi liquid behavior - a suggestive of quantum
instability in this system as expected from the above results. R.
Lortz {\it et al.} \cite{Filippov1} showed that the resistivity of
ZrB$_{12}$ exhibits linear temperature dependence in a large
temperature range of 50K - 300K. The resistivity below 50K of the
normal phase is complex and does not show $T^2$-dependence. Both,
the resistivity and specific heat data exhibit signature of
anharmonic mode of Zr-vibrations. All these observations indicate
complexity of the electronic properties, different from a typical
Fermi-liquid system consistent with the photoemission results.

A dip appears at the Fermi level at temperatures much higher than
the superconducting transition temperature reminiscent of a
pseudogap feature. In order to investigate this further, we study
the fitting parameters simulating the spectral functions in Fig. 8.
The energy gap, $\Delta$ is found to be about 7.3 meV and remains
almost unchanged in the whole temperature range studied. This gap is
significantly larger than the value predicted (1-2 meV) considering
BCS behavior. The magnitude of $\Gamma_{1}$, which is a measure of
single particle scattering rate, is quite large and found to
decrease with the decrease in temperature. On the other hand,
$\Gamma_{0}$, the inverse pair lifetime is found to be significant
and smaller than 2$\Delta$ suggesting proximity to a BCS limit for
this system. $\Gamma_0$ exhibits temperature dependence quite
similar to that of $\Gamma_1$. The decrease in $\Gamma_0$ manifests
gradual enhancement of the electron pair lifetime with the decrease
in temperature that eventually leads to the superconducting phase
upon attaining coherence among the pairs. Thus, the temperature in
the vicinity of 70 K exhibiting the onset of pair formation could be
the critical temperature, $T^*$ in this system. The observation of
the signature of an energy gap and finite electron pair lifetime
above $T_c$ is quite similar to that observed for under-doped
cuprates\cite{Randeria} and is an indication of precursor effect in
this system. Clearly, the electronic structure of this system is
complex, appears to be at the crossover of BCS and unconventional
behaviors and further studies are required to understand the origin
of pseudogap \& unusual spectral lineshape.


In summary, we studied the electronic structure of ZrB$_{12}$,
predicted to be a complex conventional superconductor employing high
resolution photoemission spectroscopy and \emph{ab initio} band
structure calculations. The valence band spectra exhibit multiple
features with dominant B 2$p$ orbital character close to the Fermi
level. The experimental results could be captured reasonably well
within the local density approximations. The electronic states close
to Fermi level have $t_{2g}$ symmetry and the filling of the bands
is in the proximity of Lifschitz transition. The spectral lineshape
near the Fermi level in the high resolution spectra exhibit
deviation from typical Fermi liquid behavior. A dip at the Fermi
level emerges above the superconducting transition temperature and
gradually becomes prominent at lower temperatures. The analysis of
the spectral functions within the phenomenological descriptions
suggests finite electron pair lifetime above $T_c$ as a signature of
a precursor effect to the superconducting transition in this system.

\section*{Method}

\subsubsection{Sample preparation and characterization}

Single crystals of ZrB$_{12}$ were grown by floating zone technique
in a Crystal Systems Incorporated (CSI) four-mirror infrared image
furnace in flowing high purity Ar gas at a pressure of 2 bar
\cite{Geetha}. The quality of the crystal was confirmed and
orientation was determined from $x$-ray Laue - diffraction images.
The magnetization measurements using a Qunatum Design MPMS
magnetometer show a sharp superconducting transition temperature,
$T_{c}$ of 6.1 K \cite{biswas_thesis}.

\subsubsection{Photoemission measurements}

The photoemission measurements were performed using a Gammadata
Scienta R4000 WAL electron analyzer and monochromatic laboratory
photon sources. $X$-ray photoemission (XP) and ultraviolet
photoemission (UP) spectroscopic measurements were carried out using
Al $K\alpha$ (1486.6 eV), He {\scriptsize II} (40.8 eV) and He
{\scriptsize I} (21.2 eV) photon energies with the energy resolution
set to 400 meV, 4 meV and 2 meV, respectively. The melt grown sample
was very hard and possesses no cleavage plane. Therefore, the sample
surface was cleaned using two independent methods; top-post
fracturing and scraping by a small grain diamond file at a vacuum
better than 3$\times$10$^{-11}$ torr. Both the procedures results to
similar spectra with no trace of impurity related signal in the
photoemission measurements. The measurements were carried out in
analyzer transmission mode at an acceptance angle of 30$^o$ and
reproducibility of all the spectra was ensured after each surface
cleaning cycle. The temperature variation down to 10 K was achieved
by an open cycle He cryostat from Advanced Research systems, USA.

\subsubsection{Band structure calculations}

The energy band structure of ZrB$_{12}$ was calculated using full
potential linearized augmented plane wave method within the local
density approximation (LDA) using Wien 2k software \cite{wien2k}.
The energy convergence was achieved using 512 $k$-points within the
first Brillouin zone. The lattice constant of 7.4075 \AA\ was used
considering the unit cell shown in Fig. 1 \cite{Kiev}. In order to
introduce electron-electron Coulomb repulsion into the calculation,
we have employed LDA+$U$ method with an effective electron
interaction strength, $U_{eff}$ ($= U - J$; $J$ = Hund's exchange
integral) setting $J$ = 0 following Anisimov et al.\cite{anisimov}.
Consideration of finite $J$ did not have significant influence on
the results. We have carried out the calculations for various values
of electron correlation upto 6 eV for Zr 4$d$ electrons and 4 eV for
B 2$p$ electrons. Spin-orbit interactions among the Zr 4$d$
electrons are also considered for the calculations.

To calculate the $x$-ray photoemission spectrum, we have multiplied
the partial density of states obtained by the band structure
calculations by the corresponding photoemission cross sections for
Al $K\alpha$ energy. These cross-section weighted PDOS is convoluted
by two Lorentzians with the energy dependent full width at half
maxima (FWHM) for the hole \& electron lifetime broadenings and a
Gaussian with FWHM representing the resolution broadening. The sum
of these spectral functions provides the representation of the
experimental valence band spectrum.

\section*{ACKNOWLEDGEMENTS}

The authors, K. M. and N. S. acknowledge financial support from the
Department of Science and Technology under the Swarnajayanti
Fellowship Programme. One of the author, G. B. wishes to acknowledge
financial support from EPSRC, UK (EP/I007210/1).

\end{document}